\documentclass[fleqn,usenatbib]{mnras}

\usepackage[T1]{fontenc}

\usepackage{enumitem}

\DeclareRobustCommand{\VAN}[3]{#2}
\let\VANthebibliography\thebibliography
\def\thebibliography{\DeclareRobustCommand{\VAN}[3]{##3}\VANthebibliography}

\usepackage{graphicx}	
\usepackage{amsmath}	
\usepackage{amssymb}	
\usepackage[dvipsnames]{xcolor}
\usepackage{physics}
\usepackage{natbib}
\usepackage{cleveref}
\usepackage{array}
\usepackage{multirow}
\usepackage{comment}
\usepackage{orcidlink}
\usepackage{bm}
\usepackage{enumitem}
\usepackage{calc}
\crefformat{section}{Section #2#1#3} 
\crefformat{subsection}{Section #2#1#3}
\crefformat{subsubsection}{Section #2#1#3}
\crefformat{equation}{equation (#2#1#3)}
\crefformat{figure}{Fig. #2#1#3}
\crefformat{table}{Table #2#1#3}
\crefformat{appendix}{Appendix #2#1#3}

\usepackage{newtxtext,newtxmath}  

\newcommand{\diff}{\mathrm{d}}

\newcommand{\LCDM}{$\Lambda$CDM}
\newcommand{\imm}{\mathrm{i}}
\newcommand{\vel}{\textbf{v}}
\newcommand{\pos}{\textbf{x}}
\newcommand{\kvec}{\textbf{k}}
\newcommand{\deltak}{\delta_{\kvec}}

\newcommand{\OK}{\Omega_{K}}

\newcommand{\oDE}{\omega_{\mathrm{DE}}}

\newcommand{\oK}{\omega_{K}}
\newcommand{\ob}{\omega_{\rm b}}
\newcommand{\oc}{\omega_{\rm c}}

\newcommand{\sotto}{\sigma_{8/h}}
\newcommand{\sdodici}{\sigma_{12}}
\newcommand{\PL}{P_{\rm L}(k, z)}
\newcommand{\Pdd}{P_{\delta\delta}}
\newcommand{\Pdt}{P_{\delta\theta}}
\newcommand{\Ptt}{P_{\theta\theta}}
\newcommand{\Gadget}{\textsc{Gadget-4}}
\newcommand{\syncp}{{\it synchronization point}}
\newcommand{\syncps}{{\it synchronization points}}
\newcommand{\shapep}{\Theta_{\rm s}}
\newcommand{\evolp}{\Theta_{\rm e}}

\newcolumntype{C}{>{\centering\arraybackslash\hspace{4pt}}c<{\hspace{4pt}}}

\setlist[itemize]{leftmargin=20pt,
        itemindent=5pt,
        labelsep=5pt,
        labelwidth=25pt,
        labelindent=!}

\title[Evolution mapping for velocity statistics]{Evolution mapping II: describing statistics of the non-linear cosmic velocity field. }

\author[M. Esposito et al.]{
Matteo Esposito$^{1}$\thanks{esposito@mpe.mpg.de}\orcidlink{0000-0001-6386-0997},
Ariel G. S\'anchez$^{1}$\orcidlink{0000-0003-1198-831X},
Julien Bel$^{2}$, and 
Andr\'es N. Ruiz$^{3,4}$\orcidlink{0000-0001-5035-4913}
\\~\\
$^{1}$Max-Planck-Institut f\"ur Extraterrestrische Physik, Postfach 1312, Giessenbachstr., D-85748 Garching, Germany \\
$^{2}$Aix Marseille Univ, Université de Toulon, CNRS, CPT, Marseille, France\\
$^{3}$Instituto de Astronomía Teórica y Experimental (CONICET-UNC), Laprida 854, X5000BGR, Córdoba, Argentina\\
$^{4}$Observatorio Astronómico, Universidad Nacional de Córdoba, Laprida 854, X5000BGR, Córdoba, Argentina
}

\date{Accepted XXX. Received YYY; in original form ZZZ}

\pubyear{202X}

\begin{document}
\label{firstpage}
\pagerange{\pageref{firstpage}--\pageref{lastpage}}
\maketitle

\begin{abstract}
We extend the evolution mapping approach, 
introduced in the first paper of this series
 to describe non-linear matter density fluctuations, to statistics of the cosmic velocity field. This 
framework classifies cosmological parameters into shape parameters, 
which determine the shape of the linear matter power spectrum, 
$\PL$, and evolution parameters, which control its amplitude at 
any redshift. Evolution mapping leverages the fact that density 
fluctuations in cosmologies with identical shape parameters but 
different evolution parameters exhibit similar non-linear evolutions 
when expressed as a function of clustering amplitude. 
We analyse a 
suite of N-body simulations sharing identical shape parameters but 
spanning a wide range of evolution parameters. Using a method for 
estimating the volume-weighted velocity field based on the Voronoi 
tesselation of simulation particles, we study the non-linear 
evolution of the velocity divergence power spectrum, $\Ptt(k)$, and 
its cross-power spectrum with the density field, $\Pdt(k)$. We 
demonstrate that the evolution mapping relation applies 
accurately to $\Ptt(k)$ and $\Pdt(k)$. While this breaks down in 
the strongly non-linear regime, 
deviations can be modelled in terms 
of differences in the suppression factor, $g(a) = D(a)/a$, similar 
to those for the density field. 
Such modelling describes the 
differences in $\Ptt(k)$ between models with the same linear 
clustering amplitude to better than 1 percent accuracy at all scales 
and redshifts considered. Evolution mapping simplifies the 
description of the cosmological dependence of non-linear density 
and velocity statistics, streamlining the sampling of large 
cosmological parameter spaces for cosmological analysis.

\end{abstract}




\section{Introduction}

The emergence of organised cosmic structures has been an important focus for cosmologists since the advent of 
large-scale surveys. The millions of observed galaxies reveal an intricate network of filaments and clusters 
that we call {\it cosmic web} \citep{Peebles}. Observations of such clustering of galaxies can provide insights 
into fundamental physics and offer a way to test our understanding of gravity and of the components that make 
up our Universe \citep[see e.g.,][]{Efstathiou2002,Eisenstein2005, Cole2005, Alam2017, Alam_2021, DESIBAO2024}. However, this can only be achieved if 
such impressive galaxy clustering surveys are accompanied by comparable theoretical models and techniques. 

N-body simulations are 
the ideal tool for
detailed studies of non-linear structure formation 
\citep{Efstathiou1985, Kravtsov1997, Teyssier2002, Springel2005, Springel2021, Potter2017, Garrison2019}. 
By numerically solving the equations of motion for a large number of particles, these simulations can model 
the evolution of cosmic structures from the early universe to the present day. N-body simulations are crucial 
for testing theoretical models against observational data, as they can account for 
nonlinear gravitational effects that are difficult to capture analytically. 
However, such accurate simulations come at the cost of high computational complexity. Large suites of 
simulations are often needed to explore different cosmologies, requiring numerous CPU hours on high-performance 
computer clusters. For this reason, exploiting our understanding of cosmology is paramount to 
aiding the exploration of wide cosmological parameter spaces.

With this in mind, \citet{Sanchez2022} pointed out that a useful degeneracy is revealed if one classifies 
cosmological parameters into two groups: shape parameters, $\shapep$, which determine the shape of the linear 
matter power spectrum, $\PL$, and evolution parameters, $\evolp$, which only affect its amplitude at any 
given redshift. With this definition, the time evolution of $\PL$ in models sharing the same shape parameters 
but different evolution parameters can be mapped from one to the other by relabelling the redshifts that 
correspond to the same linear clustering amplitude. This {\it evolution mapping} relation not only simplifies comparing different cosmological models but also accurately describes 
the non-linear matter power spectrum of the full density field \citep{Sanchez2022}.

However, the density field cannot be observed directly. In particular, galaxy positions along the line of sight can 
only be inferred from their redshift, which is also influenced by the galaxies' peculiar velocities 
\citep{Kaiser_1987}. The modelling of this effect---the so-called redshift-space distortions (RSD)---is fundamental 
for producing accurate predictions of the observational data.
Moreover, these velocities arise from 
gravitational interactions within the large-scale structure, providing additional information about the underlying 
mass distribution and the dynamics of cosmic structures \citep{Guzzo_2008}. For this reason, this paper aims to 
use evolution mapping to simplify the modelling of velocity statistics.

To achieve this, we employ the {\it Aletheia} simulations. These simulations share the same shape parameters but 
adopt different evolution parameters. For each simulation, we analyze snapshots at the redshifts at which the 
different cosmologies have the same amplitude of linear matter perturbations and, thus, also the same linear 
matter power spectrum. We show in this way how the evolution mapping relation can also be applied to statistics of 
the velocity field and, in particular, to the auto-power spectrum of the velocity divergence, $\Ptt(k)$, and its 
cross-power spectrum with the density field, $\Pdt(k)$. These are two crucial ingredients in semi-analytical recipes 
for RSD \citep{Scoccimarro_2004}.

Estimating the velocity field in N-body simulations is, however, a complex task: due to the lack of particles in under-dense 
regions, such portions of the simulation are left with little information. For this reason, different methods have 
been proposed to reconstruct the velocity field in the voids between particles 
\citep[see, e.g.,][for recent implementations]{Hahn_2015,Bel_2019,Feldbrugge2024}. In this work, we employ a modification 
of an algorithm based on the approximation that the velocity field is constant in the cells of the Voronoi 
tesselation of the N-body tracers \citep{Bernardeau_1996}.

Equipped with simulations that span a wide range of evolution parameters and an algorithm to accurately 
and efficiently reconstruct the velocity field, we show how the evolution mapping relation can also be 
applied to $\Ptt(k)$ and $\Pdt(k)$. We demonstrate how the small deviations we find from the perfect mapping can be 
attributed to differences in the growth of structure histories of the different models and can be described 
with the same recipes 
proposed in \cite{Sanchez2022} as a function of the suppression factor $g(a) = D(a)/a$. 

We begin in \cref{sec:evmap} by summarising the results of \cite{Sanchez2022} on the matter power spectrum and 
argue why these should also apply to $\Ptt(k)$ and $\Pdt(k)$. In \cref{sec:sims}, we present the Aletheia 
simulations, and in \cref{sec:methods}, we describe the method used to estimate the velocity field from them. 
In \cref{sec:results} we present and discuss our results. Finally, we give our conclusions in \cref{sec:conclusions}. 
We include in \cref{apx:G4_problem} a discussion on a problem that 
emerges
when dealing with \Gadget{} 
simulations if the user is interested in the velocities of snapshots at a very precise redshift.

\section{Evolution mapping}
\label{sec:evmap}
\subsection{The matter power spectrum}

One of the successes of the standard \LCDM{} model resides in its ability to describe the physics of cosmic 
structure formation through a relatively small number of parameters. 
Although different parameter bases are equivalent, choosing one that explicitly exhibits degeneracies in the statistics of interest (in our case, the matter power spectrum) can simplify its modelling and understanding.
With this in mind, we focus on the parameters
\begin{equation}
    \Theta = (\omega_\gamma, \omega_{\rm b}, \omega_{\rm c}, \omega_{\rm DE}, \omega_K, w_0, w_a, A_{\rm s}, n_{\rm s}).
    \label{eq:full_params}
\end{equation}
Here, $\omega_i$ represents the physical energy densities of 
radiation ($\gamma$), baryons (b), cold dark matter (c), dark 
energy (DE), and curvature ($K$), given by
\begin{equation}
    \omega_i = \frac{8 \pi G}{3H^2_{100}} \rho_i,
\end{equation}
in terms of their present-day densities $\rho_i$ and $H_{100} = 100\,{\rm km}\,{\rm s}^{-1}\,{\rm Mpc}^{-1}$. We allow for a time-dependent dark energy equation-of-state parameter, $w_\mathrm{DE}(a)$, described
in terms of the linear parametrization 
\citep{Chevallier_Polaski, Linder2003}
\begin{equation} 
    w_{\mathrm{DE}}(a) = w_0 +w_a (1-a).
    \label{eq:DE_eq_state_param}
\end{equation}
Finally, the values of $A_{\rm s}$ and $n_{\rm s}$ describe the 
amplitude and spectral index of the primordial power spectrum. 

\cite{Sanchez2022} showed that the parameters of equation~(\ref{eq:full_params}) can be classified in terms of 
their impact on the linear matter power spectrum, $\PL$, into {\it shape} 
and {\it evolution} parameters. The former set includes
\begin{equation}
    \shapep = (\omega_\gamma, \omega_{\rm b}, \omega_{\rm c}, n_{\rm s}),
    \label{eq:shape_pars}
\end{equation}
which determine the shape of the primordial power spectrum 
and the transfer function. The latter set contains 
\begin{equation}
    \evolp = (\omega_{\rm DE}, \omega_K, w_0, w_a, A_{\rm s}),
    \label{eq:evol_pars}
\end{equation}
which only affect the amplitude of $P_{\rm L}(k)$ at any given redshift $z$. 
For non-flat cosmologies, the primordial power spectrum deviates from a simple power-law on large scales. Dynamic dark energy models also show scale-dependent features in $\PL$ on super-horizon scales. However, the scales at which these effects are significant are larger than the ones considered in this work. Therefore, these parameters can be treated effectively as evolution parameters.

Since changes in the values of the evolution parameters, as well as in the redshift $z$ (through the growth factor), only affect the amplitude of $\PL$, they are all degenerate with each other. 
Therefore, the effects of these parameters can be captured by a single quantity representing the amplitude of $\PL$.
This is often given in terms of the RMS variance of the linear-theory density field at a reference scale $R$ as
\begin{equation}
    \sigma^2(R) = \frac{1}{2\pi^2} \int \diff k k^2 \PL W^2(kR),
\end{equation}
where $W^2(kR)$ is the Fourier transform of a top-hat window function at a scale $R$. We follow \cite{Sanchez20} and adopt $\sdodici\coloneqq\sigma(R=12\,\mathrm{Mpc})$ as our parameter for measuring the amplitude of $\PL$. 
With this choice, the evolution parameters will only affect $\PL$ through their effect on $\sdodici$. In other words, independently of the combination of evolution parameters and redshift, the linear matter power spectra of models with the 
same shape parameters and $\sdodici$ will be indistinguishable, that is  
\begin{equation} \label{eq:evol_mapping_lin}
    P_L(k|z, \shapep, \evolp) = P_L(k|\shapep, \sdodici(z, \shapep,\evolp)).
\end{equation}
This simple relation exposes how, for a fixed set of shape parameters, 
$\PL$ of cosmologies with different evolution parameters can be mapped to 
one another by simply relabelling the redshifts that correspond to the same 
$\sdodici$.  
Following \cite{Sanchez2022}, we will hereafter refer to \cref{eq:evol_mapping_lin} as the {\it evolution mapping} relation.

We explicitly left out the physical density of massive neutrinos, 
$\omega_\nu$,  in \cref{eq:shape_pars}.  
Massive neutrinos lead to a time-dependent suppression of power at 
small scales induced by neutrino free-streaming.
This means that the growth factor becomes scale-dependent,  
breaking the evolution-mapping relation.  
However, it is possible to extend the evolution 
mapping framework to include cosmologies with $\omega_\nu\neq 0$ by considering 
a reference cosmology without massive neutrinos and $\omega'_{\rm c} = \omega_{\rm c} + \omega_\nu$. We leave a detailed description of this 
treatment for future work (Finkbeiner et al., in prep.) 
and focus here only on cosmologies without
massive neutrinos. 

Note that the degeneracy of \cref{eq:evol_mapping_lin} cannot be described in terms of 
the traditional parameter $\sotto = \sigma(R = 8\,h^{-1}{\rm Mpc})$ and that it
breaks if the so-called Hubble units, $h^{-1}{\rm Mpc}$, are adopted. The dimensionless
Hubble parameter $h = H_0/H_{100}$ is given by the sum of all physical density parameters as
\begin{equation} \label{eq:hubble_factor}
    h^2 = \sum_i \omega_i, 
\end{equation}
and as such, it represents a mixture of shape and evolution parameters. Using quantities
that explicitly depend on $h$, such as $\sotto$ or the {\it fractional} density parameters 
$\Omega_i \coloneqq \omega_i/h^2$, obscures this degeneracy. For this reason, throughout this work, we will provide all results in Mpc units.

The perfect correspondence described by \cref{eq:evol_mapping_lin} is strictly valid only in the linear regime, as non-linear dynamics couples the evolution of different Fourier modes.
Nevertheless, since $\PL$ plays a dominant role in shaping the non-linear matter power spectrum, the degeneracy outlined in \cref{eq:evol_mapping_lin} largely carries over to the non-linear regime. This effect is especially noticeable in the framework of standard perturbation theory (SPT). If we assume that the perturbation kernels are independent of cosmological parameters —which holds for an Einstein-de-Sitter (EdS) universe and is a reasonable approximation even for non-standard cosmologies \citep{Takahashi2008,Taruya2016,Garny2021}— SPT predicts that the non-linear power spectrum depends solely on $\PL$, without explicit dependence on the cosmological parameters. 
\citet{Sanchez2022} used N-body simulations of cosmologies sharing the same shape 
parameters but differing in evolution parameters to show that the matter power spectra measured at redshifts where these cosmologies have the same value of $\sdodici$ remain extremely similar.
This provides an approximate evolution 
mapping relation 
\begin{equation} \label{eq:evol_mapping}
    P(k|z, \shapep, \evolp) \sim P(k|\shapep, \sdodici(z, \shapep,\evolp)).
\end{equation}
Deviations from this mapping are found at small scales and become larger 
as the non-linear evolution becomes more prominent, i.e. with increasing 
$\sdodici$. Such 
deviations are due to the different growth histories that the models experience and 
can be modelled in terms of the suppression factor $g(a) = D(a)/a$ and its 
derivative $g'(a)=\diff g/\diff \sdodici$. \cite{Sanchez2022} proposed a 
phenomenological recipe for predicting how each model deviates from 
a reference \LCDM{} one by Taylor expanding $P(k)$ around their value 
of $g$ and $g'$ and using the difference in these quantities to predict the 
deviations in the $P(k)$ as
\begin{equation} \label{eq:Pk_corrections}
\begin{split}
    P\left(k|g, g' \right) = P\left(k|g_0, g_0'\right) &+ \pdv{P}{g}\left(k|g_0, g_0'\right ) \cdot (g - g_0) \\
    &+ \pdv{P}{g'}\left (k|g_0, g_0'\right ) \cdot (g' - g_0'),
\end{split}
\end{equation}
where $g_0, g_0'$ indicate the values at the reference model and all quantities are evaluated at the same $\shapep$
and $\sdodici$.

The ansatz of \cref{eq:Pk_corrections} is highly effective at 
describing the differences in the non-linear matter power spectra 
of different cosmologies. 
In the rest of the paper, we will show how evolution mapping can also 
be applied to the cosmic velocity field. 
In particular, we will focus on modelling the power spectrum of 
the divergence of the velocity field and its cross-power spectrum 
with the density field, and show how \cref{eq:Pk_corrections} is also effective for 
predicting these quantities.

\subsection{The cosmic velocity field}

We will now show that \cref{eq:evol_mapping} can also be applied, with some adjustments, 
to velocity statistics.
Let us start with linear perturbation theory, in which the relation between the density 
contrast $\delta$ and the peculiar velocity field $\vel$ of a pressureless fluid is given by the continuity equation \citep[see, e.g.,][for a pedagogical derivation]{Cosmo_book}
\begin{equation} \label{eq:cont_eq_real_space}
    \pdv{\delta(\pos, t)}{t} + \frac{\nabla \cdot \vel (\pos, t)}{a} = 0,
\end{equation}
where $t$ is the cosmic time associated to a comoving observer, $\pos$ are the comoving coordinates and $\vel \coloneqq a \mathrm{d} \pos/ \mathrm{d}t$ is the peculiar velocity relative to the background expansion. 
Going to Fourier space, \cref{eq:cont_eq_real_space} can be expressed as
\begin{equation} \label{eq:cont_eq_fourier_space}
    \imm \kvec \cdot \frac{\vel_\kvec}{af(a)H(a)} = - \deltak,
\end{equation}
where the logarithmic growth rate $f(a)\coloneqq \diff \ln D(a)/\diff \ln a$ 
is the logarithmic derivative of the linear growth factor $D(a)$ with
respect to the scale factor $a$.

Let us now define a rescaled velocity field, $\Upsilon_\kvec$, and its divergence, 
$\theta_\kvec$, as
\begin{equation}
  \Upsilon_\kvec \coloneqq - \frac{\vel_\kvec}{af(a)H(a)}; \qquad  \theta_\kvec \coloneqq \imm \kvec \cdot \Upsilon_\kvec.
\end{equation}
In terms of these quantities, \cref{eq:cont_eq_fourier_space} can be conveniently expressed as
\begin{equation} \label{eq:cont_theta}
    \theta_\kvec = \deltak,
\end{equation}
and we can express the relation between the auto and cross-power spectra of $\delta_\kvec$ 
and $\theta_\kvec$ as
\begin{equation} \label{eq:lin_P_theta}
     \Ptt (k) = \Pdt (k) = \Pdd (k),
\end{equation}
where $P_{\delta\delta}(k)$ corresponds to the linear-theory matter power spectrum defined above 
as $P_{\rm L}(k)$.
Hence, if evolution mapping applies to $P_{\rm L}(k)$, it must also work for the rescaled 
velocity field $\Upsilon_\kvec$. 

It is instructive to show what this rescaled velocity field corresponds to. Let us make a change 
of variables and use the overall amplitude of clustering as a time variable, or more precisely 
$\tau \coloneqq \ln \sdodici$
\begin{equation} \label{eq:rescaled_vel}
    \dv{\pos}{\tau} = \dv{\pos}{t} \dv{t}{\tau} = \frac{\vel(t)}{af(a)H(a)} = -\Upsilon (t).
\end{equation}
This implies that this rescaled velocity field is obtained with a change of time variable \citep{Nusser_Colberg_1998}.
In other words, what will be identical at the linear level in cosmologies with the same shape parameters but different evolution parameters and evaluated at the same $\sdodici$ are not the trajectories in units of $t$ but rather those in units of the clustering amplitude, namely $\ln \sdodici$. In the following, we will always refer to the velocity field and its divergence in units of $\ln \sdodici$.

\section{The Aletheia simulations} \label{sec:sims}

\begin{table}
    \centering
    \begin{tabular}{c c}
        \hline
         Parameter & Value \\
        \hline
         $\ob$ & 0.02244 \\
         $\oc$ & 0.1206  \\
         $\oDE$ & 0.3059      \\
         $\oK$ & 0    \\
         $h$ & 0.67   \\
        \hline
    \end{tabular}
    \begin{tabular}{c c}
        \hline
         Parameter & Value \\
        \hline
         $w_{0}$ & -1 \\
         $w_a$ & 0 \\
         $n_\mathrm{s}$ & 0.97 \\
          $\sigma_{12}(z=0)$ & 0.825 \\
          $A_\mathrm{s}$ & $2.127 \times 10^{-9}$\\
        \hline
    \end{tabular}
    \caption{Cosmological parameters of our reference \LCDM{} model. The value of $h$ is obtained from \cref{eq:hubble_factor}. The value of A$_\mathrm{s}$ is set to match the chosen $\sigma_{12}(z=0)$.}
    \label{tab:LCDM_params}
\end{table}

\begin{table*}
    \centering
    \begin{tabular}{cccccccc}
            \hline
            \multirow{2}{*}{Model} 
                &\multirow{2}{*}{Definition}
              & \multicolumn{6}{c}{Redshifts of snapshots at given $\sdodici$}\\  
                 &  & $z_{IC}$ & $\sdodici=0.343$ & $\sdodici=0.499$ & $\sdodici=0.611$ & $\sdodici=0.703$ & $\sdodici=0.825$ \\
            \hline
         	Model 0 & Reference \LCDM{} (\cref{tab:LCDM_params}) & 99.0 & 2.000 & 1.000 & 0.570 & 0.300 & 0.00\\
			Model 1 & \LCDM{}, $\oDE = 0.1594$ $(h=0.55)$ & 90.6 & 1.760 & 0.859 & 0.480 & 0.248 & 0.00 \\
			Model 2 & \LCDM{}, $\oDE = 0.4811$ $(h=0.79)$ & 107.1 & 2.230 & 1.137 & 0.659 & 0.352 & 0.00 \\
			Model 3 & $w$CDM, $w_{0} = -0.85$ & 103.4 & 2.100 & 1.044 & 0.590 & 0.307 & 0.00 \\
			Model 4 & $w$CDM, $w_{0} = -1.15$ & 95.9 & 1.922 & 0.964 & 0.553 & 0.293 & 0.00 \\
			Model 5 & $w_0w_a$CDM, $w_a=-0.2$ & 97.7 & 1.972 & 0.990 & 0.566 & 0.299 & 0.00 \\
			Model 6 & $w_0w_a$CDM, $w_a=0.2$ & 100.7 & 2.031 & 1.011 & 0.574 & 0.301 & 0.00 \\
			Model 7 & Non-flat \LCDM{}, $\OK=-0.05$ ($\oDE=0.3283$) & 94.3 & 1.937 & 0.978 & 0.561 & 0.297 & 0.00 \\
			Model 8 & EdS model, $\oDE = 0.0$ $(h=0.38)$ & 78.0 & 1.402 & 0.651 & 0.349 & 0.174 & 0.00 \\
            \hline
    \end{tabular}
    \caption{Details of the Aletheia simulations. Model 0 adopts a \LCDM{} model with the cosmological parameters given in \cref{tab:LCDM_params}. All remaining models are obtained from model 0 by varying one evolution parameter as described in the definition in the second column. The parameters in parentheses represent the values of $h$ or $\oDE$ adopted to compensate for a change in a density parameter. The remaining columns list the redshift of the initial conditions and of the five snapshots obtained at the given values of $\sdodici$.}
    \label{tab:sim_params}
\end{table*}

We employed a suite of N-body simulations, the {\it Aletheia} simulations, to measure how accurately evolution mapping can describe the velocity divergence auto-power spectrum $\Ptt(k)$ and its cross-power spectrum with the density field $\Pdt(k)$ in the non-linear regime. We performed these simulations with \Gadget{} \citep{Springel2021} with $1500^3$ particles and a box side length of 1492.5 Mpc ($1000 \, h^{-1}$Mpc). 
The suite spans nine different cosmologies that share the same shape parameters but differ in their evolution parameters. The reference model is a \LCDM{} universe with Planck-like values for the cosmological parameters, shown in \cref{tab:LCDM_params}. In the rest of the models, we vary around the reference values $\oDE$, $w_0$, $w_a$ and $\oK$. We also include a model with an Einstein-de Sitter (EdS) cosmology for which $\oDE = 0$. For each model that varies a density parameter (either $\oDE$ or $\oK$), we compensate with a change of $h$ or $\oDE$. Each model has a fixed $\sdodici(z=0) = 0.825$. Hence, the values of $A_s$ in the different models are adjusted to reach the given overall amplitude of $\PL$. 
A summary of the parameters for each cosmology is shown in \cref{tab:sim_params}.

We obtained two realizations for each simulation following the paired-fixed recipe 
of \citet{paired-fixed} to suppress cosmic variance. We generated the initial conditions 
(ICs) for the simulations with \textsc{2LPTic} \citep{2LPTic}. To minimise the variance 
introduced by the generation of the ICs, we generated only one pair of boxes and varied the 
starting redshift for each simulation to match the $\sdodici$ of the given box. This is allowed 
because at the same $\sdodici$ and at the linear level, all these cosmologies exhibit the same 
density distribution. Although the same is not true for the velocity field, as shown 
in \cref{eq:rescaled_vel}, it is valid for the rescaled velocity $\Upsilon$. 
Hence, for each cosmology, we rescaled the velocities from the reference ICs by the 
ratios of the corresponding factors $af(a)H(a)$.
Finally, given that \Gadget{} takes the ICs in Hubble units as input, we also rescaled 
the positions and box sizes for the models with different values of $h$.

We produced five snapshots for each model at the redshifts at which each cosmology reaches 
a given value of $\sdodici$. This means that our snapshots are at the same $\sdodici$ 
rather than at the same redshift, and thus, all these snapshots have the same $\PL$ 
(and consequently also identical linear-theory $\Ptt(k,z), \Pdt(k,z)$, \cref{eq:lin_P_theta}).

The simulations in this work replicate the ones employed in \citet{Sanchez2022}, with some differences. 
First, we replaced the Early Dark Energy (EdE) model with the EdS one. This choice comes from the fact that the EdE simulation was performed with an effective model that only considered the impact of EdE on the Hubble function and the growth factor. More generally, EdE models can also change the shape of the transfer function. Thus, we decided to substitute that model with an EdS cosmology, which is often used as a baseline for theoretical modelling and will be useful for future applications.
Secondly, the simulations of \citet{Sanchez2022} were performed starting from fixed redshift rather than fixed $\sdodici$ as described above. 

Finally, the most important change resides in the way the \Gadget{} snapshots are produced. In particular,
the simulations in \citep{Sanchez2022} were performed with a compile-time option (OUTPUT\_NONSYNCHRONIZED\_ALLOWED) to address a limitation in how \Gadget{} produces snapshots. 
As \Gadget{} uses an adaptive time-stepping scheme, it synchronizes particles only at specific times, known as \syncps.
This means particles in different regions of the simulation (dense or underdense) are updated at different rates based on their local conditions. As a result, particles are only synchronized at these designated points.
When the user requests a snapshot at a given redshift, \Gadget{}’s default behaviour is to save it at the closest available \syncp. This may result in small shifts in redshift, undesirable in our analysis in which we want to compare simulations at the exact same value of $\sdodici$.
The OUTPUT\_NONSYNCHRONIZED\_ALLOWED option allows \Gadget{} to output snapshots at the exact redshift requested by linearly drifting particle positions to the desired time. However, while the positions are adjusted to the correct redshift, the particle velocities are left unchanged, meaning they correspond to different times. As a result, the snapshot contains inconsistent velocities at different redshifts, which can introduce errors in velocity-related analyses.

To address this problem while ensuring that the snapshots are exactly at the desired redshifts, we performed this new set of simulations with a {\it start-stop} method. This method leverages the fact that all particles are naturally synchronized at the start and end of each simulation run. By stopping the simulation at each required snapshot and restarting it with adjusted start and end times, we ensure that all particles are fully synchronized at the exact redshift requested. This guarantees consistent outputs for both positions and velocities. Readers interested in a more detailed discussion of the problem can find it in \cref{apx:G4_problem}, along with tests of the {\it start-stop} method.

\section{Estimating the density and velocity power spectra} \label{sec:methods}

Estimating velocity statistics in N-body simulations (and for similar reasons in observations) is intrinsically difficult because particles trace the density field and not the velocity one.
The lack of particles in underdense regions 
is indicative of their density but can give little to no 
information on the velocity field inside them. 
Low-density regions suffer from large Poisson errors and in regions where 
there are no particles at all the velocity field remains completely undetermined
\citep[see, e.g.][]{Juszkiewicz1995,Bernardeau_1996,BernardeauEtal1997}.
 For this 
reason, different techniques have been developed for recovering the volume-weighted velocity 
field from N-body simulations 
\citep{W&B1998, RW2007,P&S2009,  Jennings2011, Jennings2012, JenningsBaughHatt2015, Zheng2013, 
KodaEtal2014, ZhengZhangJing2015, ZhangZhengJing2015, Yu2015, Bel_2019}.
A common choice is to employ the Delaunay tesselation field estimator algorithm
\citep[DTFE, ][]{Schaap2000, Cautun2011}, in which the Delaunay 
tesselation\footnote{In a Delaunay tesselation, particles are linked to each other, forming 
a set of tetrahedra such that no particle is inside the circumsphere of any 
tetrahedron, optimizing for tetrahedra to be as close to equiangular as possible.}
of the N-body particles is used to interpolate the velocity 
to every point in space under the assumption that the field is continuous and with a 
constant gradient inside each tetrahedra. This results in a
volume-weighted estimate of the velocity field instead of the mass-weighted 
outcome obtained using standard particle-in-cell techniques.

In this work, we follow a different approach based on the work of \cite{Bernardeau_1996}, 
that implies a lower-order approximation of $\vel$.  
We assume that the velocity field is constant inside each Voronoi cell\footnote{In a Voronoi tesselation, space is divided into regions around a set of 
particles such that each region contains all the points closer to its corresponding particle
than to any other, forming a set of convex polyhedra.} of the particles 
in the simulation. 
With this piece-wise approximation, the field is discontinuous at the edges of each cell 
and some small-scale information is lost. 
However, we propose a Monte-Carlo variation of this Voronoi method that compensates a small 
loss in accuracy by considerably reducing the number of calculations needed.
This approach can be implemented without calculating the full Voronoi tesselation of 
our simulations. We simply populate the boxes with 
points sampled from a distribution that evenly covers the volume (in our case, a glass-like distribution, see below),
and 
assign to each of them the velocity of the closest N-body particle. 
This results in a new particle distribution that reproduces our piece-wise approximation 
of the velocity field. 
The now 
evenly
distributed samples can be used to estimate 
the smoothed velocity field by means of commonly used mass assignment schemes.  
Hereafter, we refer to this approach as the Monte-Carlo Voronoi method (MC-Voronoi).

In our implementation, we start by populating the simulation box with a sample of particles following a 
glass-like distribution generated using the recipe of \citet{Davila-Kurban2021}. 
These samples provide a more homogeneous covering of the volume than a Poisson distribution 
with the same number of points, reducing the shot-noise contribution in our estimates. 
Each point in this distribution is assigned a velocity matching that of its closest neighbour 
among the simulation particles. 
We then use this new set of points to compute the CIC smoothed velocity field, $\Bar{\bf v}$, 
on a regular mesh 
\begin{equation}
    \Bar{\bf v}_i = \frac{\sum_j w({\bf x}_i,{\bf x}_j){\bf v}_j}{\sum_j w({\bf x}_i,{\bf x}_j)},
\end{equation}
where $w({\bf x}_i,{\bf x}_j)$ represents the CIC kernel evaluated at the positions of the 
node $i$ of the grid and point $j$ in the glass-like distribution. Note that, in contrast with directly applying a CIC assignment scheme to the N-body particles, in this case, our tracers sample the simulation box uniformly, and the velocity field thus obtained is not weighted by the density field; moreover, this method does not leave empty cells as long as the density of glass particles is high enough compared to the chosen grid size.

Other than the choice of the mass-assignment scheme, the other two important parameters 
in this procedure are the size of the grid and the number of glass-like points used to 
sample the simulation volume. In 
this work, we adopt $1024^3$ grid points to resolve the scales we are interested in. To save computation time (which grows linearly with the number of points in the sample), we only employed the same number of glass points as of particles in the simulation. 
Increasing this number can improve the accuracy of this method; however, our tests have shown that the gain is minimal in our case.

\begin{figure*}
    \centering
    \includegraphics[width=0.9\textwidth]{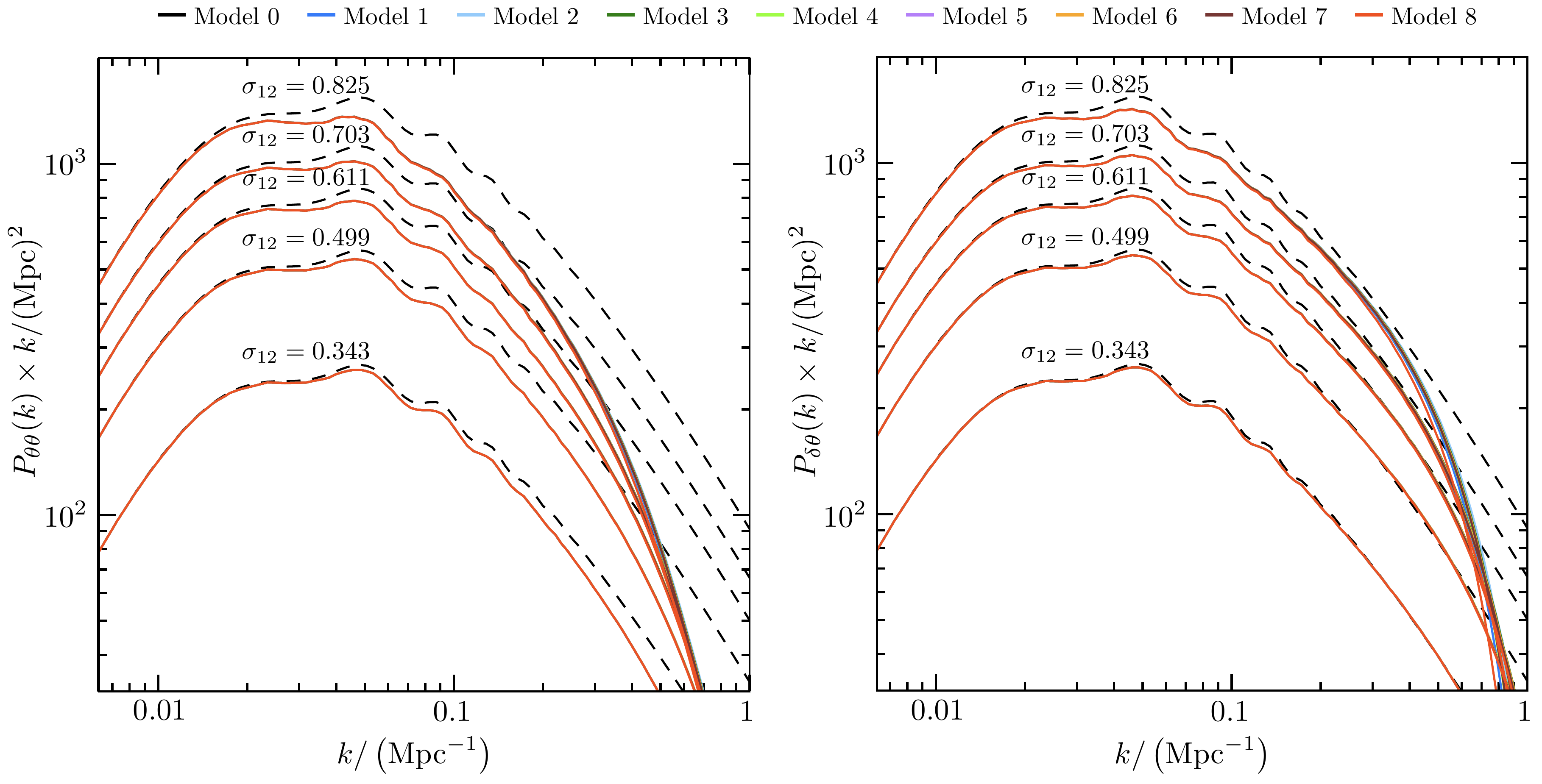}
    \caption{Velocity divergence auto- (left panel) and cross- (right panel) power spectra measured from the Aletheia simulations. The power spectra are multiplied by $k$ for a clearer display of the lines. Each colour indicates a different cosmology for which the power spectra from all 5 snapshots are shown. Rather than being at the same $z$, each snapshot is extracted at the $z$ at which each model reaches a given $\sdodici$ (indicated above each set of lines).}
    \label{fig:full_Pks}
\end{figure*}

Once the velocity field has been computed on a regular grid, it can be used to estimate 
various statistics. In this work, we are interested in the velocity divergence auto power 
spectrum and its cross-power spectrum with the density field. 
We also use the CIC assignment scheme to reconstruct the density field on the same grid of $1024^3$ points. We evaluated a second mesh grid shifted by half a grid size for each snapshot to correct for aliasing through an interlacing technique \citep{Sefusatti2016}.
For calculating the power spectra, we used the publicly available library Pylians\footnote{\url{https://pylians3.readthedocs.io/en/master/}} \citep{Pylians}, which Fourier-transforms 
the velocity field, obtains $\theta_\kvec$ in Fourier space, and uses it to calculate 
$\Ptt(k)$ and $\Pdt(k)$.

Although more accurate methods for estimating the smoothed velocity field exist in 
the literature, we chose this procedure due to its computational simplicity. Since we are 
interested in differences and ratios between power spectra rather than their absolute values, 
the MC-Voronoi method employed in this work is enough to reach solid conclusions. 
In particular, we performed preliminary tests comparing our results with the method proposed in \cite{Bel_2019}, which employs a Delaunay field estimator and found perfectly compatible outcomes. 
We leave a detailed comparison between these different methods for future work (Esposito et al., in prep). 

\section{Results} \label{sec:results}

In this section, we present the power spectra estimated with the technique described in \cref{sec:methods}. \cref{fig:full_Pks} shows the $\Ptt(k)$ and $\Pdt(k)$ estimated from the Aletheia simulations in the left and right panels, respectively. The solid lines correspond to the results for the cosmologies described in \cref{tab:sim_params}, 
while the dashed lines indicate the linear theory 
predictions. As expected, given that the snapshots of the 
different cosmologies correspond to the same values of 
$\sdodici$, all models have the same power spectra at 
large scales but present some small deviations at small 
scales where the evolution mapping relation becomes 
approximate. The measurements of $\Pdt(k)$ show a steep 
drop and a change of sign (which is not visible in the 
figure) that correspond to the appearance of vorticity. 
While the velocity field has no vorticity at the linear 
level, it is produced by the non-linear evolution of the 
density fluctuations \citep{P&S2009, Hahn_2015}. Such 
vorticity disrupts the cosmic flow and breaks the tight 
correlation between the density and velocity fields. 
Hence, vorticity causes the cross-power spectrum to 
drop at the scales when it becomes important and then 
change its sign at the scale of shell crossing.

\begin{figure*}
    \centering
    \includegraphics[width=0.9\textwidth]{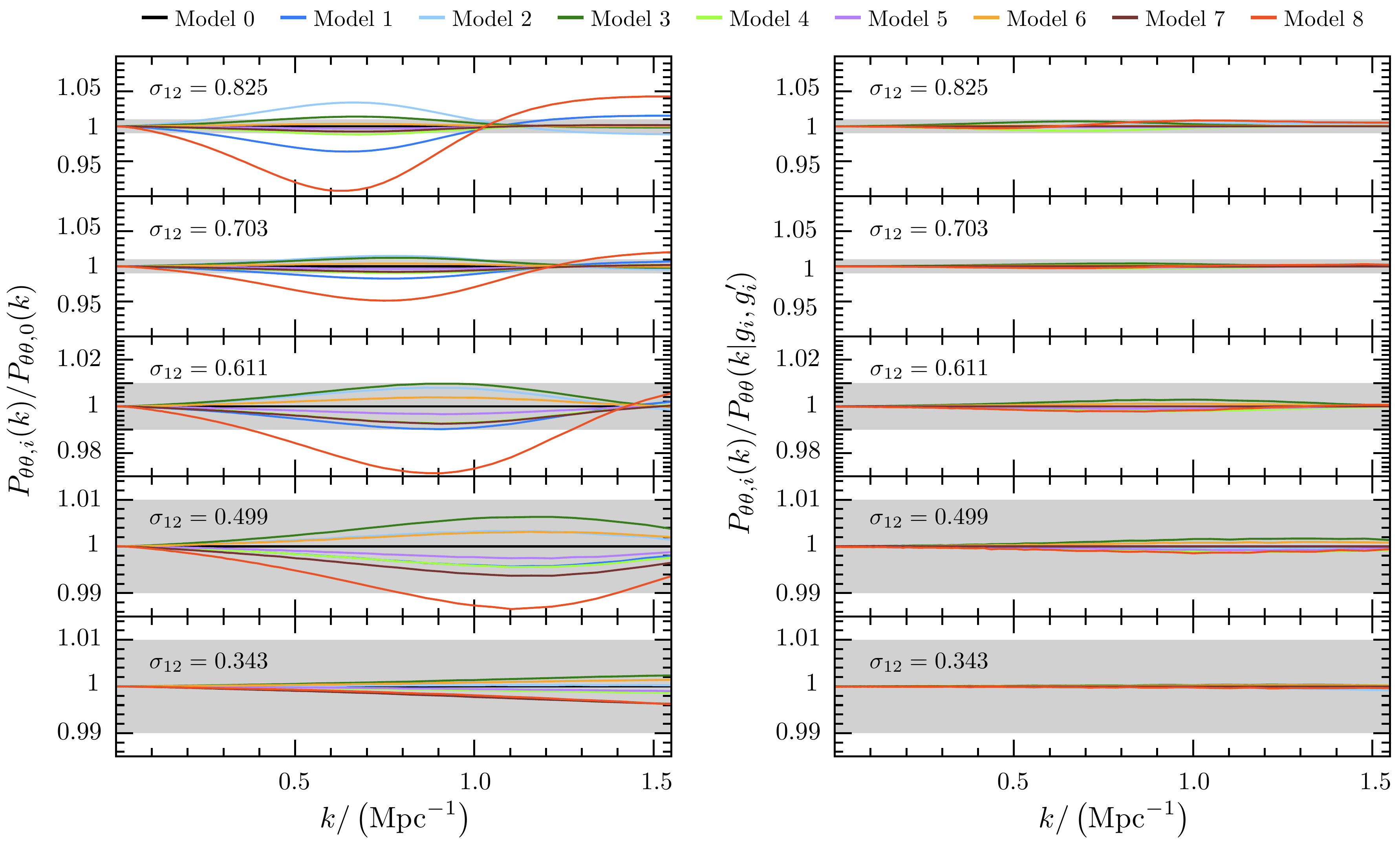}
    \caption{Ratios of velocity divergence auto-power spectra measured from the Aletheia simulations to that of model 0 (left panel) and to the predictions of \cref{eq:Pk_corrections} (right panel). The different colours correspond to different cosmologies described in 
    \cref{tab:sim_params}. From bottom to top, the panels show the five snapshots extracted from the simulations at increasing values of $\sdodici$. The shaded grey regions show 1 per cent deviations from model 0.}
    \label{fig:Vel_Pks_ratios}
\end{figure*}

The left panel of \cref{fig:Vel_Pks_ratios} shows the 
ratios of the measurements of $\Ptt(k)$ in the different models to the one measured in the reference $\Lambda$CDM cosmology (model 0). 
The different panels correspond to our five reference values of $\sdodici$ increasing from bottom to top and 
the colour-coding matches that of \cref{fig:full_Pks}. 
The deviations from model 0 become more significant 
with increasing $\sdodici$, corresponding to 
increasing levels of non-linearities. 
Going from larger to smaller scales, we can identify 
different regimes when comparing the different models to 
our reference one: (1) at large linear scales, the $\Ptt$ 
are almost undistinguishable because they share the same 
linear theory power spectra; (2) going to smaller scales, 
evolution mapping breaks down, and the different growth 
histories cause the power spectra to diverge from each 
other in the same direction as the $\Pdd(k)$ do; (3) with 
the appearance of vorticity and as more particles 
shell-cross, the sign of the correlation between 
$\theta$ and $\delta$ changes \citep{Hahn_2015} and 
thus, the trend in the deviations between the different 
models is reverted and grows in the opposite direction. 
The maximum deviations in the range of scales that can 
be measured from these simulations go from less than 
0.5 per cent at $\sdodici = 0.343$ to approximately 5 
per cent at $\sdodici=0.825$. The exception is model 8, 
which reaches $10$ per cent deviations. This 
last case corresponds to an EdS cosmology with an 
extreme value for $h=0.378$ and thus exhibits much 
larger deviations than the other models.

As shown by \citet{Sanchez2022}, the differences in the 
matter power spectra of these models are related to 
their different structure-formation histories, which can be described by the differences in the growth 
suppression factors $g(\sdodici)$ and its derivative 
$g'(\sdodici)$. We show here that this is also valid 
for velocity statistics and that 
we can use \cref{eq:Pk_corrections} not only for predicting $\Pdd(k)$, but also $\Pdt(k)$ and $\Ptt(k)$ 
based on the results of our reference $\Lambda$CDM model. 

We estimated the derivatives in \cref{eq:Pk_corrections} by assuming that this relation is exact for models 1 and 7 and 
solving the system of equations for $\partial P(k)/\partial g$ and $\partial P(k)/\partial g'$. We used \cref{eq:Pk_corrections} to approximate the measurements of $\Ptt(k)$ of the remaining models.

The right panel of \cref{fig:Vel_Pks_ratios} shows the ratios of the $\Ptt(k)$ measured from the simulations to the ones estimated using this recipe\footnote{To see the results for the $\Pdd(k)$, check \cite{Sanchez2022}}.
To simplify the comparison with the 
original model differences, the ranges 
of both axes are the same as those of 
the left panel. 
Equation (\ref{eq:Pk_corrections}) can indeed capture the differences between 
model 0 and the other cosmologies. The maximum residual differences range from approximately 
$0.1$ per cent at $\sdodici = 0.343$ 
to per cent level deviations at 
$\sdodici=0.825$. 
The $\Ptt(k)$ of the EdS cosmology (model 8), which in the left panel shows the largest differences, 
is also recovered with the same level of accuracy as the other models. 

\begin{figure*}
    \centering
    \includegraphics[width=0.9\textwidth]{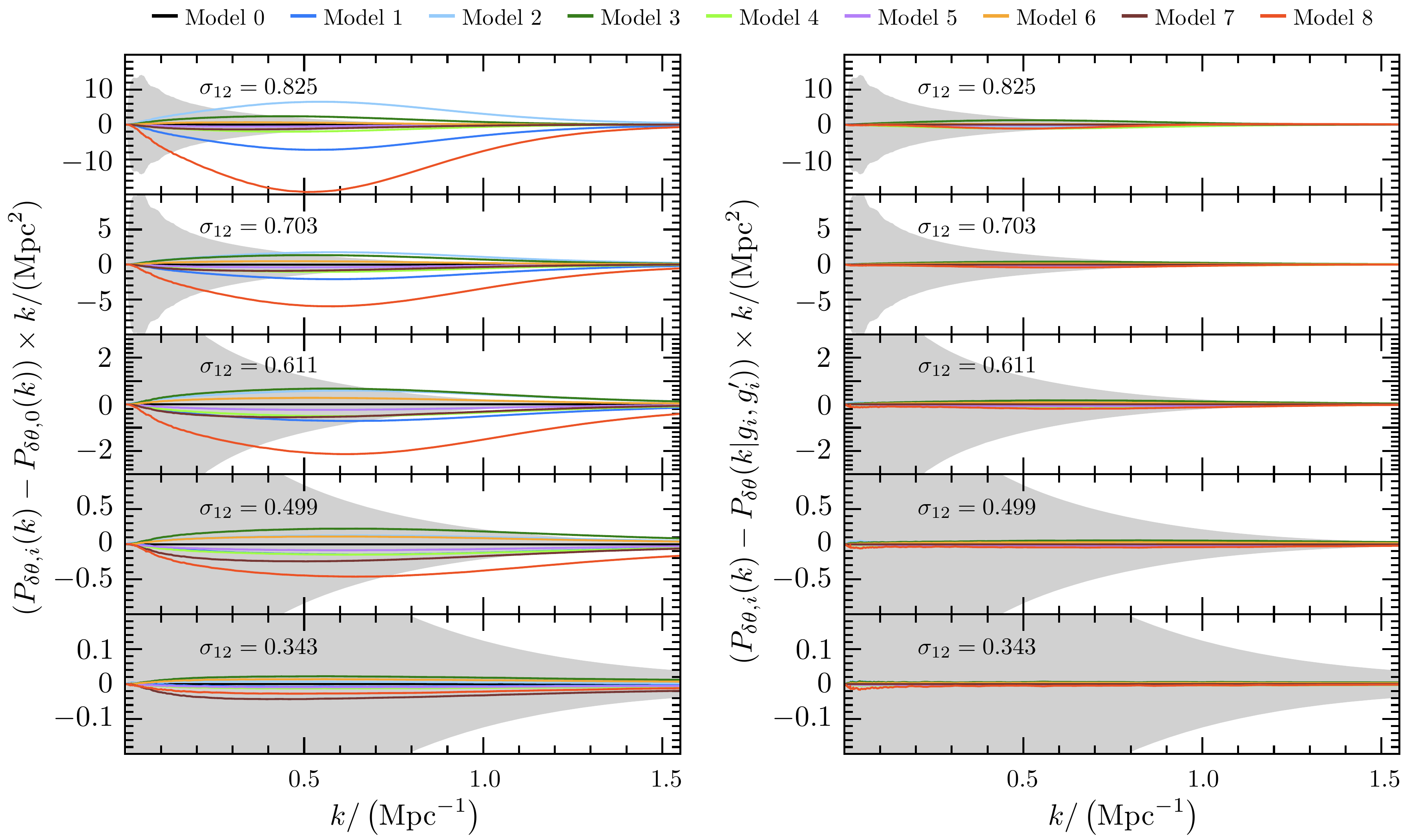}
    \caption{Differences of cross-power spectra of the velocity divergence with the density field measured from the Aletheia simulations with respect to model 0 (left panel) and to the predictions of \cref{eq:Pk_corrections} (right panel). The power spectra are multiplied by $k$ for a clearer display of the lines. The different colours correspond to different cosmologies. From bottom to top, the panels show the five snapshots extracted from the simulations at increasing values of $\sdodici$. The shaded grey regions show values within $\pm 1$ per cent of the power spectra of model 0.}
    \label{fig:Cross_Pks_diff}
\end{figure*}

\cref{fig:Cross_Pks_diff} shows the 
result of the same procedure applied to 
$\Pdt(k)$. The left panels show the differences between $k \times \Pdt$ measured from the simulations and the one of model 0. The 
right panels show the differences between the same measurements and the predictions obtained using the ansatz of \cref{eq:Pk_corrections}. 
As mentioned before, the non-linear $\Pdt$ crosses zero at the scale of shell crossing at which the correlation between the density and velocity fields changes sign. This scale is reached by our measurements for some of the snapshots. Hence, we decided to show absolute differences in this case instead of relative differences, which are undefined at the zero-crossing.
In order to discuss these results similarly to the ones presented in \cref{fig:Vel_Pks_ratios}, we include in each panel a shaded region that includes values within $\pm 1$ per cent of $k \times \Pdt$ of model 0. We also notice in this case that the lines leave the $1$ per cent region as we move to smaller scales (larger values of $k$) and larger clustering (larger values of $\sdodici$).

The right panel of \cref{fig:Cross_Pks_diff} shows how \cref{eq:Pk_corrections} can predict also $\Pdt(k)$ with high accuracy, with differences with the measured power spectra that are reduced well within the grey shaded regions.
Note that \cref{eq:Pk_corrections} does not include ratios of power spectra, and it is thus well-defined at all scales. 
However, it is more difficult to give a meaningful estimate of the (relative) error that one would make when using \cref{eq:Pk_corrections} to predict the power spectra of the other models.

There is a deep connection between the 
evolution mapping approach and the 
cosmology-rescaling method proposed by 
\citet{Angulo_2010}, which adapts a 
cosmological simulation to match a different 
target cosmology. When the cosmologies share 
the same shape parameters, this rescaling 
involves identifying the redshift at which the 
amplitude of the linearly-evolved matter 
density fluctuations (i.e., the value of $
\sigma_{12}$) in the target cosmology matches 
that of the original cosmology. At this point, 
velocities are rescaled by the ratio of the 
factors $af(a)H(a)$ appearing in the continuity 
equation, aligning with the methodology 
employed in our work. The correction for 
differences in the growth histories in terms 
of $g(\sigma_{12})$, as given in 
\cref{eq:Pk_corrections}, is analogous to the 
rescaling of the concentration-mass relation 
proposed by \citep{Contreras2020}. The 
simulation rescaling method has demonstrated 
practical success in various studies 
\citep[e.g.,][]{Angulo_2010,Angulo2021,Ruiz2011,Zennaro2023}. 
Our evolution mapping framework can be seen as a 
theoretical basis for why this approach 
works. Specifically, the spatial and 
temporal scaling that best matches 
the shape and amplitude of the linear power 
spectra between the two cosmologies also 
results in a good agreement of the full non-
linear structure formation.

Our results show that the mapping of \cref{eq:Pk_corrections} indeed applies also to the velocity 
field and, therefore, describes the full particle trajectories. 
Thus, 
any other statistic extracted from N-body simulations would follow, to some extent, the evolution 
mapping relation of \cref{eq:evol_mapping}.

By adopting physical density parameters and absolute Mpc units, we can take advantage of the 
evolution-mapping degeneracy in the cosmological parameters. This is useful for increasing the 
accuracy and performance of emulators and fitting functions,  but most importantly, it simplifies 
our understanding of structure formation.
This is yet another reason for avoiding the use of the traditional Hubble units, which 
obscure the impact of the different cosmological parameters on the evolution of cosmic 
structure.

\section{Conclusions} \label{sec:conclusions}

We followed the approach of evolution mapping proposed in \cite{Sanchez2022} and applied 
it to velocity statistics. In particular, we employed a suite of dark matter only N-body simulations 
that share the same shape parameters (which determine the shape of $\PL$), but cover a wide 
range of evolution parameter values (which determine the redshift evolution of the amplitude of 
$\PL$) to show how---when evaluated at the same $\sdodici$---the statistics of the velocity field, 
namely $\Ptt(k)$ and $\Pdt(k)$, are remarkably similar up to few percent differences at non-linear scales. 
The deviations appearing at smaller scales are due to the different growth histories that these 
models experience. Thus, we showed how modelling these deviations in terms of $g$ and $g'$, can 
predict the $\Ptt(k)$ and $\Pdt(k)$ of a given model with sub-percent to percent level accuracy.

Measuring the smoothed velocity field using simulation particles as tracers is a challenging 
task, due to the lack of information in underdense regions. In this work, we adopted a new 
approach for estimating the velocity field on a grid through a Monte-Carlo sampling of the field 
under the assumption of a piece-wise constant approximation in the cells of a Voronoi tesselation. 
Given that a comparison of the power spectra of our different cosmologies shows little sensitivity to the specific method adopted, to infer them, we leave more detailed tests and 
comparisons of this and other methods for future works (Esposito et al. , in preparation).

We highlight a limitation in \Gadget{} when performing velocity-related analyses with simulation snapshots that require a specific redshift. The OUTPUT\_NONSYNCHRONIZED\_ALLOWED option, while providing accurate particle positions at exact redshifts, leads to inconsistencies in velocities because particles are not synchronized in time, introducing potential biases that are difficult to predict. We thus recommend using a {\it start-stop} method instead, which we discuss in more detail in \cref{apx:G4_problem} where we address the problem.

 As discussed in \citet{Sanchez2022}, evolution mapping 
 can streamline the design of emulators for the 
 non-linear power spectrum and other density field 
 statistics by reducing the number of required 
 simulation nodes and redshift outputs, thus lowering 
 computational demands. Traditional emulators often 
 struggle with high-dimensional cosmological parameter 
 spaces and assume restrictive parameter ranges. 
 Evolution mapping simplifies this process by using a 
 reference set of fixed evolution parameters and sampling 
 over the parameter space defined by shape parameters 
 and $\sigma_{12}$. Each node in this space corresponds 
 to a simulation at a single redshift where the desired 
 $\sigma_{12}$ is achieved. The emulator's predictions 
 can then be adapted to any desired cosmology using the 
 recipe in \cref{eq:Pk_corrections}. This approach not 
 only reduces the dimensionality of the parameter space 
 but also enhances prediction accuracy. Our results 
 indicate that the same method can be extended to build 
 emulators for non-linear velocity field statistics and 
 the impact of redshift-space distortions on other 
 clustering statistics. This extension further broadens 
 the applicability of evolution mapping, making it a 
 versatile tool for constructing comprehensive models of 
 cosmological observations.

By adopting physical density parameters 
and absolute Mpc units, we 
can exploit the evolution-
mapping degeneracy in 
cosmological parameters more 
effectively. This approach not 
only enhances the accuracy and 
performance of emulators and 
fitting functions but also 
significantly simplifies our 
understanding of structure 
formation. Using Mpc units 
avoids the complications 
associated with traditional 
Hubble units, which obscure the 
impact of different cosmological 
parameters on the evolution of 
cosmic structures. This clarity 
provides a compelling reason to 
standardize measurements in 
absolute units, simplifying the 
interpretation of cosmological 
analyses.

\section*{Acknowledgements}
We would like to thank Raul Angulo and the BACCO project for providing useful scripts for benchmarking the codes used in this work.
We also would like to thank 
Sofia Contarini, 
Carlos Correa, 
Andrea Fiorilli, 
Luca Fiorino, 
Nelson Padilla, 
Alejandro Perez, 
Andrea Pezzotta, 
Agne Semenaite
and Matteo Zennaro
for their help and useful discussions. 
This work was 
funded by the Deutsche Forschungsgemeinschaft (DFG, German Research Foundation) under Germany´s Excellence Strategy – EXC 2094 – 390783311.
The Aletheia simulations were carried out and post-processed on the HPC system 
Cobra of the 
Max Planck Computing and Data Facility (MPCDF) in Garching, Germany. 
We acknowledge support from the European Research Executive Agency 
HORIZON-MSCA-2021-SE-01 Research and Innovation programme under the 
Marie Sklodowska-Curie grant agreement number 101086388 (LACEGAL).

\section*{Data Availability}

The data underlying this article will be shared with the corresponding author at a reasonable request.



\bibliographystyle{mnras}
\bibliography{bibliography} 




\appendix

\section{Problems with non-synchronized outputs in Gadget4} \label{apx:G4_problem}

In this appendix, we show how non-synchronized velocities can bias the velocity statistics in Gadget simulations in a way which is difficult to predict. To integrate particle trajectories most efficiently, {\sc Gadget} (in all its versions) employs an adaptive time step scheme. In this way, particles in dense regions have smaller time steps than those in voids. The code can thus accurately integrate complicated trajectories for the former without wasting time with multiple time steps for straight trajectories of the latter. This implies that particles might be non-synchronised at a given time in the simulations, i.e. the equation of motions of different particles has been integrated up to different times. However, the time domain decomposition in {\sc Gadget} ensures that, through their evolution, particles get synchronized at the so-called \syncps. Since this set of points is discrete, the user will often request snapshots at redshifts at which particles are not fully synchronized. 

The default option in \Gadget{} 
for these scenarios is to output a snapshot at the closest \syncp. This can ensure that the output is fully consistent at the cost of small perturbations in the redshift of the output. This approach works in most scenarios but becomes problematic when the user needs a snapshot at a specific redshift. This can be the case, for example, when the user is comparing different cosmologies at the exact same redshift. Or, in the case of the Aletheia simulations, at the exact same $\sdodici$. 

One possible way to overcome this problem is to use the compile-time option OUTPUT\_NONSYNCHRONIZED\_ALLOWED (hereafter ONSA). In this case, when the code reaches the time of a requested snapshot, it linearly interpolates the particle positions to the given time (i.e. it does a final drift) by using the lastly updated velocities while leaving the latter untouched. This solution is exact in predicting the particle positions. Still, it introduces an unpredictable error in the velocities: the snapshot will contain velocities that not only are not at the requested redshift but not even all at the same redshift. The user can, in principle, overcome this problem by reducing the largest time step allowed, but how effective this will be is difficult to estimate a priori.

We noticed this is the default behaviour of some versions of {\sc Gadget-3}. Still, given that this version was never released publicly and branched into many versions, we refrain from making any tests with it in this paper. However, even though the effect is subtle and only affects velocities, we wish to make users who might overlook this issue consider it when using simulations for velocity statistics (e.g. analyses of velocity power spectra or redshift-space quantities).

As a quick solution that does not require code modifications, we employed a {\it start-stop} method in our simulations that we suggest applying when the user needs the output at a specific redshift. Given that the time domain decomposition is performed from the {\it TimeBegin} to the {\it TimeEnd} of the simulation, all the particles will be synchronized in all cases at these points. Hence, if the user performs the simulations until the first snapshot, then restarts from there until the second and so on, while each time changing {\it TimeBegin} and {\it TimeEnd}, this will ensure that all snapshots always have fully synchronized particles. The suggested approach for re-starting a simulation in \Gadget{} is using restart files. However, this approach is intended not to break the time integration scheme, saving the time domain decomposition calculated at the beginning of the simulation in the restart files. In this way, restarting a simulation at what was saved to be its end time will result in improper behaviour. As an alternative, we suggest saving the simulation snapshots in double precision when needed for a restart. This will ensure minimum information loss through the restarting of the simulation.

\begin{figure}
    \centering
    \includegraphics[width=0.9\columnwidth]{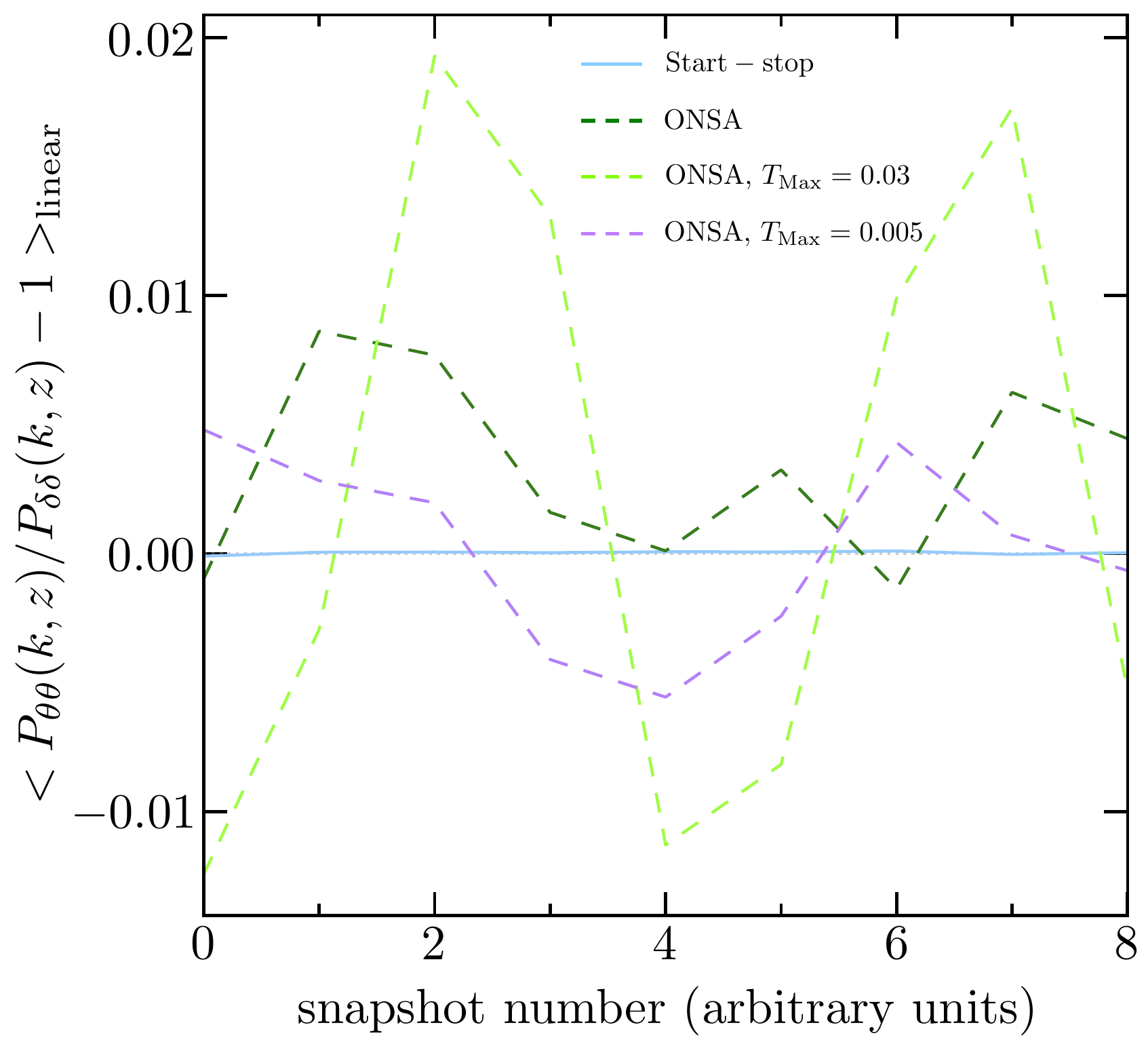}
    \caption{Relative differences in the amplitude of $\Ptt$ and $\Pdd$ across the simulation snapshots. The ratios between power spectra are calculated as the mean of $\Ptt/\Pdd$ in the first 8 bins of the measurements. The solid line represents the results obtained with the start-stop method, while the dashed lines are obtained using the ONSA flag. Where not indicated, the maximum allowed time step $T_\mathrm{Max}$ is the default one (0.01)}
    \label{fig:synchro-mess}
\end{figure}

To show the effect of non-synchronized velocities, we performed a series of N-body simulations with different maximum allowed time steps. We employed the public versions of \Gadget{}. All simulations include $512^3$ dark matter particles in a (2000 Mpc/$h)^3$ cubic volume. We perform each simulation in two ways: with the option of non-synchronized outputs and with the {\it start-stop} method. We estimate the error introduced in the velocity field (in particular in its divergence) by comparing $\Ptt$ and $\Pdd$ on linear scales; using non-synchronized velocities also changes the non-linear part of the $\Ptt$, but here, we focus our comparisons on scales for which we have solid theoretical predictions (see \cref{eq:lin_P_theta}). To obtain sensible results at these scales, we reduce the cosmic variance by starting our simulations from fixed-pair initial conditions at $z=49$ with the 2LPTic code. We obtain 9 snapshots at $z=\{20, 15, 10, 8, 5, 4, 3, 2, 1\}$.
We show in \cref{fig:synchro-mess} the relative differences in the amplitude of the $\Ptt$ and $\Pdd$, estimated as the mean value of the power spectra in the first 8 bins. We present these measurements as a function of snapshot number to show how these deviations do not show any obvious correlation with redshift, being only determined by the choice of output redshift requested by the user. We present results for 3 different values of the maximum time step ($T_\mathrm{Max}$) a particle may take (in $\ln(a)$): 0.01 (the default value), 0.03 and 0.005. We note that decreasing the maximum time step decreases the maximum error in the amplitude of $\Ptt$. Nonetheless, the behaviour at a given snapshot is difficult to predict and reaching very high precision (like the one we needed for this work) would require very small $T_\mathrm{Max}$, which largely increases the computation time. We thus encourage users who need accurate measurements of the velocity field to rather use the start-stop method suggested here (solid line in \cref{fig:synchro-mess}), which correctly captures the expected amplitude of $\Ptt$.


\bsp	
\label{lastpage}
\end{document}